\documentclass[aps,prb,twocolumn,floatfix]{revtex4}

\usepackage{amsmath}
\usepackage{amssymb}
\usepackage{multirow}
\usepackage{graphicx}
\usepackage{rotating}
\usepackage{verbatim}
\usepackage{longtable}

\begin{document}

\title{Spin-phonon induced magnetic order in Kagome ice}

\author{F.A.\ G\'omez Albarrac\'{\i}n}
\affiliation{IFLP - Departamento de F\'{\i}sica, Universidad Nacional de La Plata, C.C. 67, 1900 La Plata, Argentina}

\author{D.C.\ Cabra}
\affiliation{IFLP - Departamento de F\'{\i}sica, Universidad Nacional de La Plata, C.C. 67, 1900 La Plata, Argentina}

\author{H.D.\ Rosales}
\affiliation{IFLP - Departamento de F\'{\i}sica, Universidad Nacional de La Plata, C.C. 67, 1900 La Plata, Argentina}

\author{G.L.\ Rossini}
\affiliation{IFLP - Departamento de F\'{\i}sica, Universidad Nacional de La Plata, C.C. 67, 1900 La Plata, Argentina}

\begin{abstract}
We study the effects of lattice deformations on the  Kagome spin ice, with Ising spins coupled by nearest neighbor exchange
and long range dipolar interactions, in the presence of in-plane magnetic fields.
We describe the lattice energy according to the Einstein model, where each site distortion is treated independently.
Upon integration of lattice degrees of freedom, effective quadratic spin interactions arise.
Classical MonteCarlo simulations are performed on the resulting model, retaining up to third neighbor interactions, under different
directions of the magnetic field.
We find that, as the effect of the deformation is increased, a rich plateau structure appears in the magnetization curves.
\end{abstract}

\maketitle

\section{Introduction}

Spin ice systems  \cite{SpinIceScience2001} have been the object of intense study in the last couple of decades.  These materials are an experimental
evidence of high magnetic frustration, showing a residual low temperature entropy and magnetic disorder.  Some  compound
examples are  Ho$_2$Ti$_2$O$_7$ \cite{SpinIce1997,SpinIce1998} and Dy$_2$Ti$_2$O$_7$ \cite{SpinIceDisp1,SpinIceDisp2}, where the rare earth ions form a lattice of corner sharing tetrahedra (pyrochlore lattice). The magnetic degrees of freedom have large spins, and therefore
can be described semiclassically. Due to intense crystal fields, spin directions are locally forced to point either in or out
from the tetrahedra centers, meaning that the spins can be modeled by local Ising variables.
The interplay between antiferromagnetic nearest neighbor exchange couplings and strong dipolar interactions in these materials leads to a highly
frustrated and degenerate ground state satisfying the so-called ice rules:\cite{IceRulesBernal,Pauling}
$2$ spins in and $2$ spins out of each tetrahedron.

The magnetization curves of the pyrochlore systems under an external magnetic field in the $\langle 111
\rangle$ direction show a well known plateau at $1/3$ of saturation.\cite{Shastry2002,MatsuhiraHiroi2002}
Above this plateau the apical spins are completely aligned with the external magnetic field, 
which suggests that the relevant physics stems from the transverse Kagome layers.
This paper is motivated by recent experiments \cite{GrigeraExper} which have identified a plateau-like feature above $1/3$ magnetization,
when the magnetic field is slightly tilted with respect to the  $\langle 111 \rangle$ direction.
This feature cannot be interpreted within the standard model for pyrochlore spin ice \cite{SpinIceScience2001}, nor with the inclusion of further neighbor exchange couplings\cite{Yavorskii2008} and
careful treatment of the long range dipolar couplings.\cite{Chufo}
Motivated by the need to understand the physics above the $1/3$ plateau, we have considered the influence of the phonon degrees of freedom in the magnetic properties of the pyrochlore. 
One should recall that for large spins the effects of the phonons appear generically enhanced by an $S^2$ factor.

In the high field regime (above the pyrochlore $1/3$ plateau) the apical spins are aligned with the field and the remaining physics could be described, 
in a first approximation, by decoupled Kagome planes. In the present work we consider a Kagome ice model under the influence of a magnetic field along the plane,
which mimics the tilting of the magnetic field in the experiments. We include the effects of phonons, which induce a rich plateau structure.

The paper is organized as follows: In Section \ref{model} we introduce the Kagome ice model for the effective description of pyrochlore spins in active layers in the regime of interest, 
including nearest neighbor exchange and dipolar interactions, and the spin interactions induced by lattice fluctuations.
In Section \ref{results} we present magnetization curves under in-plane magnetic fields, obtained by simulated annealing.
Several plateaux and their magnetic ordered structure are described. Section \ref{conclusions} is devoted to discussion and conclusions. 

\section{Kagome ice model}\label{model}

\begin{figure} [!ht]
\includegraphics[width=0.45\textwidth]{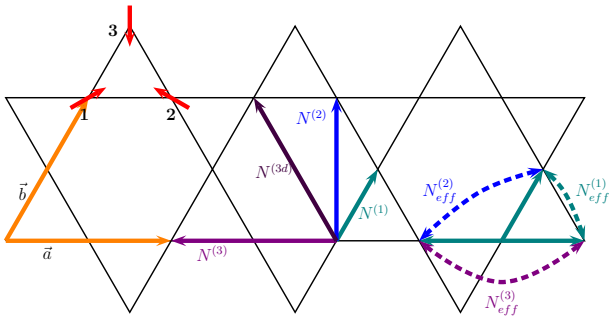}
\caption{
\label{fig:NN}
Kagome lattice. In red, the local reference directions for Ising spins on each site
($\sigma_i=+1$ is referred to as "in"). First, second and third nearest neighbors are indicated.
On the right, effective interactions arising from  $\vec{F}^{(1)}_{ij}$ (green arrows) and bridged by a site $i$ are shown by dashed lines.
}
\end{figure}

We are interested in the description of Ising pyrochlore systems in the regime where apical spins magnetization is saturated.
As mentioned in the introduction, once the apical spins of the tetrahedra are aligned with the external magnetic field, the
remaining spins lie on Kagome planes.
As a first step, we do not consider the out-of (Kagome) plane components of the spins and consider
a Kagome ice planar model,\cite{KagomeIce} {\it i.e.} local spin directions point towards or outwards the center of the
triangles of the Kagome unit cell.
Each site $i$ allocates a local Ising spin $\vec{S}_i = S\sigma_i\breve{e}_i$, where $S$ is
the  spin magnitude, $\sigma_i= \pm 1$  is the Ising variable (+1 being ``in'' and -1 ``out'') and $\breve{e}_i$ is the local
reference direction ($1$, $2$, $3$ in Fig.\ \ref{fig:NN}).
This simplified model may be also relevant to artificial permalloy arrays with Kagome geometry .\cite{Moller-Moessner-2009}
The standard model for pyrochlore spin ice\cite{SpinIceScience2001} includes exchange
antiferromagnetic interactions only for nearest neighbors ($N^{(1)}$) and long range dipolar interactions;
the Hamiltonian on the regular Kagome lattice then reads
\begin{eqnarray}
H_0 &=& J_0\sum_{\langle ij \rangle^{(1)}}\vec{S}_i\cdot\vec{S}_j \nonumber\\
&+& Da^3\sum_{i \not= j}\left[\frac{\vec{S}_i\cdot\vec{S}_j}{(r_{ij}^0)^3}-\frac{3(\vec{S}_i\cdot\breve{r}_{ij})
(\vec{S}_j\cdot\breve{r}_{ij})}{(r_{ij}^0)^3}\right] \nonumber\\
&-&\vec{h} \cdot \sum_{i}\vec{S}_{i}
\label{H_0}
\end{eqnarray}
where $J_0$ is the antiferromagnetic $N^{(1)}$ exchange interaction coupling, $D$ is the strength of the dipolar coupling,
$a$ the distance between nearest neighbors, $r_{ij}^0$ the distance between any pair of spins at sites $i$ and $j$
and $\breve{r}_{ij}$ is the unit vector from site $i$ to site $j$.
$\vec{h} = h_x\breve{x} +h_y\breve{y}$ is the external magnetic field in the Kagome plane, $\breve{x}$
being perpendicular to one of the spin directions (say $3$ in Fig.\ \ref{fig:NN}) and $\breve{y}$ parallel to it.
In the pyrochlore setting, $\vec{h}$ takes into account the deviation of the magnetic field with respect to the
$\langle 111 \rangle$ direction. The out of plane component of the spins could be easily included, together with the  $\langle 111 \rangle$
component of the magnetic field. This generalization together with its relation to magnetization experiments in
dysprosium pyrochlore \cite{SpinIceDisp1,SpinIceDisp2} will be considered in future work.

In order to include the effects of lattice deformations in the magnetic order,
we introduce deformations $\vec{u}_i$ in the site positions
\begin{equation}
 \vec{r}_i =  \vec{r}^{0}_i + \vec{u}_i,
\end{equation}
so that the distance between sites $i$ and $j$ is distorted from the regular lattice. At first order
\begin{equation}
r_{ij}=|\vec{r}_j - \vec{r}_i| \approx  r_{ij}^0 + \breve{r}_{ij}\cdot (\vec{u}_j-\vec{u}_i).
\end{equation}
The effect of the deformations in the exchange interaction is taken at linear order to be
\begin{equation}
J(r_{ij}) \approx J_0\left[1-\alpha\,\breve{r}_{ij}\cdot (\vec{u}_j-\vec{u}_i)\right]
\end{equation}
where $\alpha = -\frac{1}{J_0}\left.\frac{\partial J}{\partial r_{ij}}\right|_{r_{ij}=r^{0}_{ij}} > 0$
is the spin-phonon coupling constant.
Correspondingly, corrections to dipolar interactions  are considered at first order by varying distances
in the second line of Eq.\ (\ref{H_0}).

We treat the elastic degrees of freedom in the adiabatic limit, assuming large ion masses which is appropriate in the case of Dy$_2$Ti$_2$O$_7$.
There are different models to describe the energy cost of lattice deformations.\cite{BergmanBalentsPyrochloreBonds}
One of them is the bond phonon model,\cite{BondsPenc} describing acoustic modes, where the elastic energy depends on bond length deformation
but each bond is allowed to independently expand or contract (ignoring geometrical constraints),
{\em i.e.} variables $\delta \vec{r}_{ij}=\vec{u}_j-\vec{u}_i$ are independent. We have explored the effect of these modes in the effective description  and we trivially observe that they just lead to a constant shift in the energy.
The other standard choice is the Einstein phonon model \cite{KagomePhononsWang} describing optical modes, where the elastic energy is quadratic on each site displacement and truly independent deformations $\vec{u}_i$ can be exactly integrated. The spin-phonon Hamiltonian reads
\begin{equation}
H = H_{0} + \sum_{i}\left(\frac{K}{2}(\vec{u}_{i})^2 + \vec{u}_i \cdot \sum_{j \neq i} \vec{F}_{ij}\right).
\end{equation}
Here $\vec{F}_{ij}$ collects all terms proportional to $\vec{u}_i$ and containing $\sigma_i \sigma_j$,
arising from a first order expansion of the variation of $H_0$ with lattice distortions (see the explicit expressions below).
In this case, phonon degrees of freedom
are easily integrated to yield an effective Hamiltonian for the magnetic degrees of freedom at a given temperature.
In order to discuss the validity of this integration and subsequent approximations,
we set $J_0$ as the energy scale and $a$ as the length scale to introduce dimensionless parameters
$d = \frac{D}{J_0}$ for the dipolar interaction strength,
$k = \frac{Ka^2}{J_0}$ for the phonon stiffness and
$\lambda = a\alpha$ for the linear spin-phonon coupling.

The standard Gaussian integration over elastic thermal fluctuations in the presence of linear interactions assumes
that both the width and mean of the thermal distribution of displacements given by $e^{-\beta H}$ are much smaller than the lattice distance $a$.
This requires for the phonon stiffness that $k\gg \frac{k_B T}{J_0}$,
which is valid at low enough temperature $T$, and that the interaction factors $\vec{F}_{ij}$ satisfy
\begin{equation}
\left|\sum_{j \neq i} \vec{F}_{ij}\right|\ll kJ_0.
\label{condition}
\end{equation}
The explicit expression for $\vec{F}_{ij}$ at first neighbors includes the spin-phonon coupling and a dipolar term
\begin{equation}
 \vec{F}_{ij^{(1)}}=\frac{J_0 S^2}{a}\left(-\frac{\lambda}{2}+\frac{21}{4}d\right)\breve{r}_{ij}\sigma_i \sigma_j,
\label{F1}
\end{equation}
while the longer range $\vec{F}_{ij^{(n)}}$ \footnote{In standard notation $ij^{(1)} \equiv <i,j>$, $ij^{(2)} \equiv <<i,j>>$, etc.} only include dipolar terms and decay with distance as $1/r_{ij}^4$.
In particular, for second neighbors one finds
\begin{equation}
 \vec{F}_{ij^{(2)}}=-\frac{J_0 S^2}{a}\frac{5}{12}d\,\breve{r}_{ij}\sigma_i \sigma_j
\label{F2}
\end{equation}
and for third neighbors the numerical factor decays to $3/16$. Taking $\vec{F}_{ij^{(1)}}$ as the significative contribution,
Eq.\ (\ref{condition}) requires
\begin{equation}
S^2 \left|-\frac{\lambda}{2}+\frac{21}{4}d\right|\ll k
\label{condition-F1}
\end{equation}

After Gaussian integration the effective Hamiltonian reads
\begin{equation}
H_{eff}=H_0 - \sum_{i} \frac{a^2}{2kJ_0}\sum_{j,k \in N(i)} \vec{F}_{ij}\cdot\vec{F}_{ik},
\label{H_eff}
\end{equation}
where $N(i)$ in the summation in the last term refers to neighbors of each site $i$.
Notice that $\vec{F}_{ij}\cdot\vec{F}_{ik}$ contains $\sigma_i^2=1$, so it is proportional to $\sigma_j \sigma_k$;
\footnote{In this sense we have stated that a bond phonon model leads to a trivial effective
Hamiltonian where all Ising variables are squared.}
thus effective corrections to interactions between Ising spins at sites $j$, $k$,
arise from the summation of $\vec{F}_{ij}\cdot\vec{F}_{ik}$ terms in Eq.\ (\ref{H_eff}) over all $i\neq j,k$.
We find it convenient to depict each contribution to $j$, $k$ interactions as bridged by a site $i$.
In this sense, notice that when both factors $\vec{F}_{ij}$ and $\vec{F}_{ik}$ refer to nearest neighbors,
the site $i$ bridges interactions between first, second and third neighbors
as shown in Fig.\ \ref{fig:NN}. Longer range factors $\vec{F}_{ij^{(n)}}$ give rise to increasingly long range effective interactions, with smaller couplings.

In order to tailor a tractable effective model, we proceed to truncate the range of neighbors $N(i)$
in Eq.\ (\ref{H_eff}).
To this aim we compare in detail the effective corrections to first neighbors effective Ising coupling
arising only from $\vec{F}_{ij^{(1)}}$
with those including second range factors $\vec{F}_{ij^{(2)}}$.
The relative weight of such second range corrections is negligible if
\begin{equation}
\frac{5\sqrt{3}}{6} d \ll \left|\frac{21}{4}d-\frac{1}{2}\lambda\right|
\label{F2_vs_F1}
\end{equation}
Longer range contributions from $\vec{F}_{ij^{(n)}}$ are even smaller, due to the dipolar decay.
Condition (\ref{F2_vs_F1}) is largely satisfied in the case of dysprosium pyrochlore, where $\lambda$ can be
estimated to be of order $20$,\cite{zhou2011hig} whereas values of $d \approx 1/3$\cite{SpinIceScience2001} as we consider below. 
Thus we neglect  $\vec{F}_{ij^{(n)}}$ for $n\geq 2$ in what follows.

Regarding the original long range dipolar interactions in $H_0$, it is known that a truncation is more
sensible in the Kagome lattice than in the pyrochlore lattice, because of lower dimensionality.\cite{Moller-Moessner-2009}
According to the range of effective corrections kept, we also truncate long range  dipolar interactions retaining up to third neighbors.

The truncated effective hamiltonian finally reads
\begin{multline}
\frac{H_{eff}}{J_0 S^2} =    \left[ J^{(1)}_{eff}\sum_{\langle ij \rangle^{(1)}} \sigma_i\sigma_j +
J^{(2)}_{eff}\sum_{\langle ij \rangle^{(2)}} \sigma_i\sigma_j \right. \\
+\left.
J^{(3)}_{eff}\sum_{\langle ij \rangle^{(3)}} \sigma_i\sigma_j +
J^{(3d)}_{eff} \sum_{\langle ij \rangle^{(3d)}} \sigma_i\sigma_j \right]  \\
 - \sum_i \sigma_i\left[ \tilde{h}_x (\breve{e}_i)_x +
 \tilde{h}_y (\breve{e}_i)_y\right]
 \label{Heff}
\end{multline}
where $\langle ij \rangle^{(1)}$, $\langle ij \rangle^{(2)}$, $\langle ij \rangle^{(3)}$ and $\langle ij \rangle^{(3d)}$
refer to first $N^{(1)}$, second $N^{(2)}$, third along triangle edges $N^{(3)}$, and third along hexagon diagonals
$N^{(3d)}$ neighbors respectively, as depicted in Fig.\ \ref{fig:NN}. The dimensionless magnetic field $\tilde{h}_\alpha$ is defined as $h_\alpha/(J_0S)$, with $\alpha =x,y$.
The effective couplings are given by
\begin{eqnarray}
J^{(1)}_{eff}&=&-\frac{1}{2}+\frac{7}{4}d - \delta\nonumber\\
J^{(2)}_{eff}&=&-\frac{5}{12\sqrt{3}}d+\delta\nonumber\\
J^{(3d)}_{eff}&=&\frac{1}{8}d\nonumber\\
J^{(3)}_{eff}&=&-\frac{5}{32}d + 2\delta\nonumber\\
\label{J's-effectives}
\end{eqnarray}
with $\delta=\frac{S^2}{4k}\left(-\frac{\lambda}{2}+\frac{21}{4}d\right)^2$. 
In a realistic pyrochlore setting, the corrections to dipolar interactions at second and third neighbors generated by phonon degrees of freedom could be related to
those included as exchange interactions in Ref.~\onlinecite{Yavorskii2008}, for tuning diffuse elastic neutron scattering data in dysprosium titanate.

\section{Monte Carlo simulations - Results and Discussion}\label{results}

We analyze the effective Hamiltonian in Eq.~(\ref{Heff}) in a regime where exchange and first order dipolar interactions compete yielding a frustrating Ising interaction ($J^{(1)}_{eff}>0$), and where the effective
parameters $J^{(1)}_{eff}$ and $J^{(2)}_{eff}$ have the same order of magnitude (strong frustration regime). To this aim we consider $d\simeq 1/3$ for the rest of the paper.
For this values of $d$, all considered interactions indeed compete.
It should be noticed that $J^{(1)}_{eff}$ is positive up to $\delta=0.06$, thus favoring frustration, and $J^{(2)}_{eff}$ remains negative up to $\delta \approx 0.08$ (see Eq.~(\ref{J's-effectives})).

We performed Monte Carlo simulations for Kagome lattices of $N = 3\times L^2$ sites, with  $L = 12, 18, \ldots , 36$, by  conventional single-spin flip \cite{MCLandau} plus implementation of a
tempering algorithm (annealing technique),\cite{Kirkpatrick1983} lowering the temperature in a $T_{i+1}=T_i\times 0.9$ scheme, down to lowest $T=0.0042 J_0 S^2/k_B$.
At every magnetic field and temperature we discarded $2\times 10^{6}$ Monte Carlo steps (MCS) for initial relaxation and data were collected during subsequent $4\times 10^{6}$ MCS. Monte Carlo runs for the same parameters with different seeds gave no significant variations, thus no error bars are reported in average magnetization data.

We focus on the low temperature phase diagram, as a function of the single parameter $\delta$ in the presence of an external magnetic field.
To determine the different phases we computed the normalized magnetization under magnetic fields applied along the $\breve{x}$ and $\breve{y}$ direction, defined as
\begin{equation}
M_{\alpha}=\frac{1}{M_{\alpha,s}}\sum_{i=1}^{N}(\vec{S}_i)_\alpha
\end{equation}
where $\alpha=x,y$ and $M_{\alpha,s}$ is the saturation magnetization along $\alpha$.

\subsection{Magnetic field parallel to one of the spins}

\begin{figure}[h!]
\centering
\includegraphics[width=0.4\textwidth]{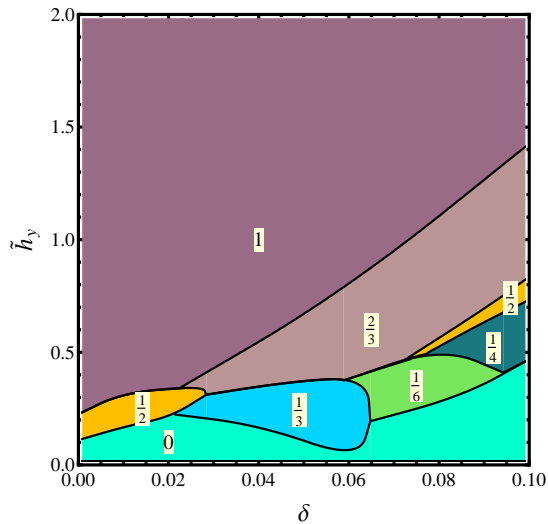}
\caption{\label{fig:phasediagram_magy} Phase diagram $h_y$ vs.\ $\delta$  for a $3\times L^2$ ($L=30$) sites Kagome lattice.
The numbers indicate the normalized magnetization of the several plateaux.}
\end{figure}

We first study the equilibrium average magnetization $\langle M_y\rangle$ under fields $\vec{h} = h_y\breve{y}$, from zero to saturation.
We performed simulations for $d=0.32$ and  $\delta$ from $0$ to $0.1$,  where the condition  in Eq.\ (\ref{condition}) is satisfied, in steps of $0.001$.
The resulting phase diagram is shown in Fig.\  \ref{fig:phasediagram_magy}.

\begin{figure}[htb]
\centering
\includegraphics[width=0.48\textwidth]{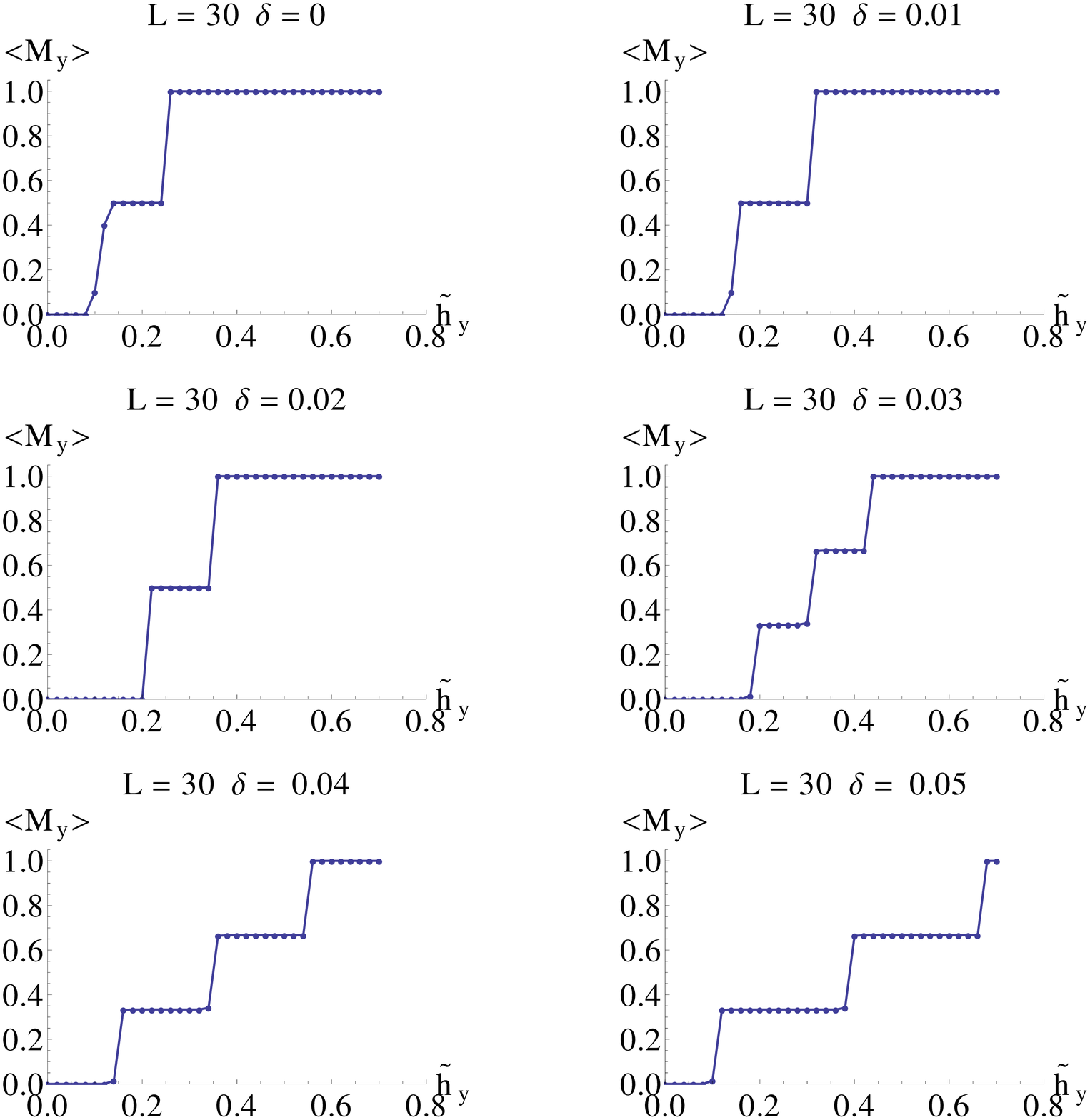}
\caption{
\label{fig:magL30} Average magnetization per spin vs.\ $h_y$ for  a $3\times L^2$ ($L=30$) sites Kagome lattice.
}
\end{figure}

We find that for $\delta=0$, which corresponds to phonon stiffness $k\to\infty$ (no lattice deformations),
there are two plateaux, for  $\langle M_y\rangle =0$ and $\langle M_y\rangle =1/2$ (Fig.\ \ref{fig:magL30}a-\ref{fig:magL30}c).
These plateaux, present by the only effect of dipolar interactions,
are stable lowering the value of $k$ (allowing for lattice deformations) up to  $\delta\simeq0.02$ where  $J^{(1)}_{eff}\simeq J^{(3d)}_{eff}\simeq 2|J_{2}^{eff}|/3$
and $J^{(3)}_{eff}\simeq 0$ changes sign from negative to positive adding frustration to the system.

For larger $\delta$ ($> 0.03$), there is a transition: the $1/2$ plateau ``splits'' into two
plateaux at $\langle M_y\rangle =1/3$ and $\langle M_y\rangle =2/3$, which widen as $\delta$ increases (Fig.\ \ref{fig:magL30}d-\ref{fig:magL30}f).
Close to $\delta=0.06$, $J^{(1)}_{eff} \simeq 0$ and the effective system is dominated by the third neighbors couplings with $J^{(3d)}_{eff}\simeq J^{(3)}_{eff}/2$ and $|J^{(2)}_{eff}|< J^{(3)}_{eff}/4$.
We recall that for  $\delta > 0.06$, $J^{(1)}_{eff}$ is negative and does not favor frustration, therefore the lattice configurations no longer satisfy the Kagome ice rules.
In this region, where the system forms  frustrated antiferromagnetic sublattices coupled by $J^{(3d)}_{eff}$ and $J^{(3)}_{eff}$,
with smaller $J^{(1)}_{eff},J^{(2)}_{eff}$, a series of plateaux appear: $\langle M_y\rangle=1/6$ for $0.065<\delta<0.085$, and $\langle M_y\rangle=1/4$ and $\langle M_y\rangle=1/2$ again for $0.085<\delta$.

\begin{figure}[htb]
\includegraphics[width=0.45\textwidth]{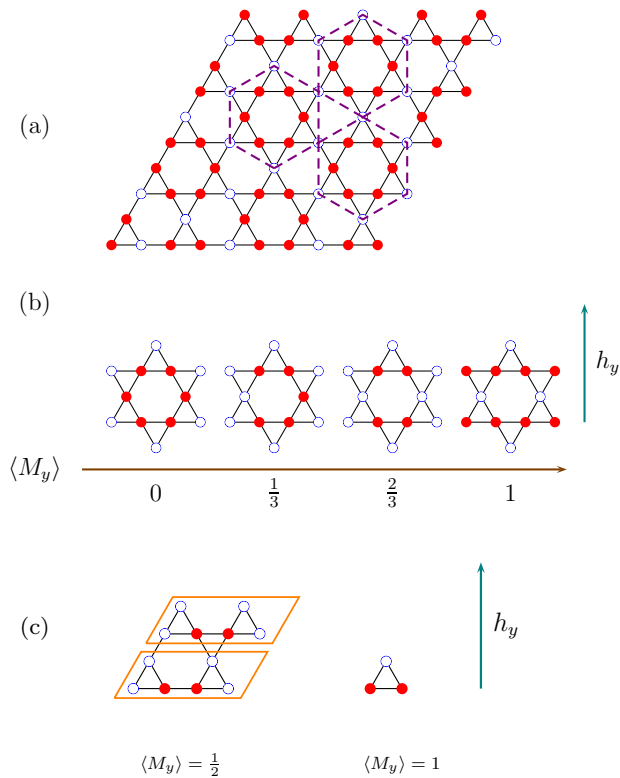}
\caption{\label{fig:spin-structure-1}
Magnetic unit cells in the Kagome lattice in different magnetization plateaux for $h_x=0$.
Open blue circles indicate $\sigma_i=-1$ (out) and full red circles $\sigma_i=+1$ (in).
(a): $\langle M_y\rangle=0$ tiling and its magnetic unit cell;
(b): Magnetic unit cell for $\langle M_y\rangle=0,1/3,2/3$ and $1$.
At the $1/3$ plateau both the figure shown or its specular reflection can be found;
(c): $\langle M_y\rangle=0$  and $1$.
For the last one the magnetic unit cell showing in (b) consist of only three sites.
}
\end{figure}

\begin{figure}[htb]	
\includegraphics[width=0.45\textwidth]{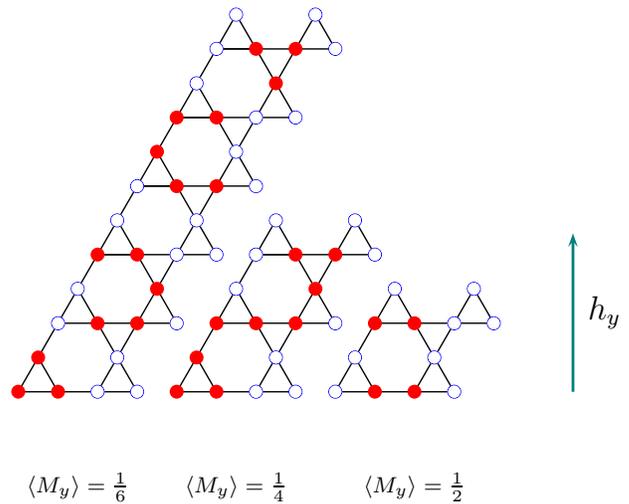}
\caption{\label{fig:spin-structure-2}
Magnetic unit cells in the Kagome lattice in different magnetization plateaux in the large $\delta$ regime. The $1/6$, $1/3$ and $1/2$ plateaux at the large $\delta$ regime.
}
\end{figure}

The local spin configurations at the different plateaux show magnetic order, which we describe by repetition of magnetic unit cells.
These are sketched for each plateau in Figs.\ \ref{fig:spin-structure-1} and \ref{fig:spin-structure-2}.

At $\langle M_y\rangle=0$, $\langle M_y\rangle=1/3$ and $\langle M_y\rangle=2/3$
the magnetic pattern is obtained by translations along the $\{\vec{a}+\vec{b},2\vec{a}-\vec{b}\}$ vectors
(depicted in Fig.\  \ref{fig:NN})
of a $9$-sites magnetic unit cell consisting of three Kagome unit cell triangles (sites $1$, $2$, $3$) forming a star.
In Fig.\ \ref{fig:spin-structure-1}a we show in detail such tiling at $\langle M_y\rangle=0$, where the $\sigma_i=+1$ are
arranged in closed hexagons inside each star, surrounded by $\sigma_i=-1$.
In  Fig.\ \ref{fig:spin-structure-1}b the content of the magnetic unit cell is shown for the following plateaux:
at $\langle M_y\rangle=1/3$ one of the spins in a site $3$ of every hexagon flips, so the $\sigma_i=+1$ form a ``C'',
while in the $\langle M_y\rangle=2/3$ plateau another $\sigma_3=+1$ flips  to $-1$.
At this point, every spin in every site $3$ is aligned with the external magnetic field.
Finally, the lattice reaches its saturation configuration where the star  consists of three identical triangles.

In contrast to the plateaux described before by $9$-sites magnetic unit cell, the rest of the plateaux exhibit different
size magnetic unit cells.
In the $\langle M_y\rangle=1/2$ regime shown in Fig.\ \ref{fig:phasediagram_magy} for small $\delta$,
the $6$-sites magnetic unit cell is  a combination of two reflected Kagome unit cell triangles with
$\sigma_3=-1$ and $\sigma_1=-\sigma_2$, as shown in Fig.\ \ref{fig:spin-structure-1}c.

We recall that all of the plateaux mentioned above have lattice triangles satisfying the Kagome ice rules.
However, for larger values of $\delta$ the magnetic configurations break the Kagome ice rules and we find:
at $\langle M_y\rangle=1/6$ there is a $36$-sites magnetic unit cell,
at $\langle M_y\rangle=1/4$ a $18$-sites magnetic unit cell
and finally  at $\langle M_y\rangle=1/2$ a $12$-sites magnetic unit cell.
All these magnetic unit cells are shown in Fig. \ \ref{fig:spin-structure-2}.

It is interesting to notice that for any of the plateaux shown in Fig.\ \ref{fig:phasediagram_magy}
the $\mathbb{Z}_2$ reflection symmetry of the system in presence of $h_y$  is not broken ($\langle M_x\rangle=0$).

\subsection{Magnetic field perpendicular to one of the spins}

\begin{figure}[!ht]
\centering
\includegraphics[width=0.4\textwidth]{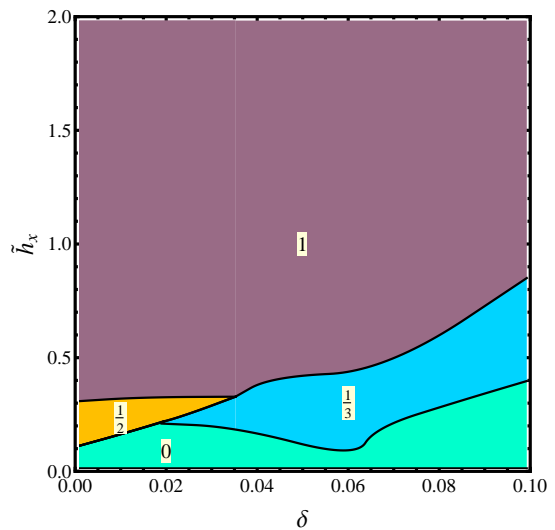}
\caption{\label{fig:phasediagram_magx} Phase diagram $h_x$ vs.\ $\delta$. The numbers show the normalized magnetization of the plateaux.
Only the $\langle M_x\rangle=1/3$ plateau is induced by lattice deformations.}
\end{figure}

\begin{figure}[tb]
\includegraphics[width=0.48\textwidth]{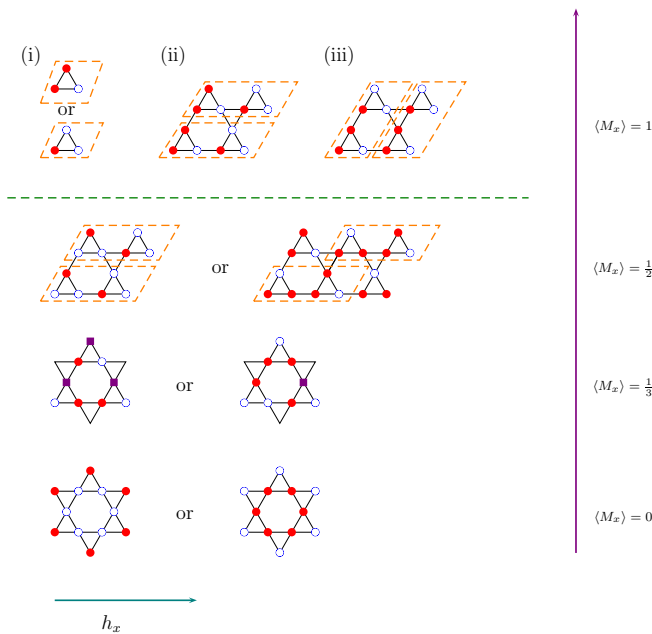}
\caption{\label{fig:spin-structure-3} Magnetic unit cells in different magnetization plateaux for $h_y=0$ in the small $\delta$ regime. Going  from bottom to top, we show magnetic orders at the various plateaux present in the magnetic phase diagram shown in Fig.\ \ref{fig:phasediagram_magx}. Purple squares indicate random spin orientation.}
\end{figure}

We proceeded in the same way for $\vec{h} = h_x\breve{x}$, ranging from zero to $\langle M_x\rangle$ saturation.
The main difference with respect to the previous case is that the spin at each site $3$ is perpendicular to
the magnetic field and is not affected by the Zeeman coupling.
The resulting magnetic phase diagram is shown in Fig.\ \ref{fig:phasediagram_magx}.
For $\delta=0$, as in the case above,
there is a plateau at $\langle M_x\rangle=1/2$ extending up to $\delta\lesssim 0.038$
where the effective model is dominated by $|J^{eff}_2|\simeq J^{eff}_{3d}$ with $J^{eff}_1\simeq J^{eff}_3\simeq J^{eff}_{3d}/2$.
At this plateau we find two possible $6$-sites magnetic unit cells, both with no net magnetization in the $\breve{y}$ direction.
The magnetic pattern is built by translation along $\{2\vec{a},\vec{b}\}$ or $\{2\vec{a},\vec{a}+\vec{b}\}$,
as shown in the second line in Fig.\ \ref{fig:spin-structure-3}.

\begin{figure}[tb]
\includegraphics[width=0.4\textwidth]{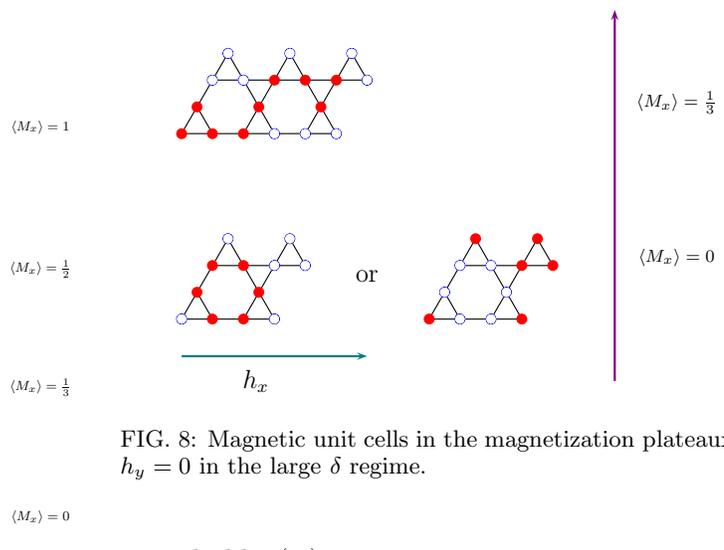}
\caption{\label{fig:spin-structure-4}
Magnetic unit cells in the magnetization plateaux for $h_y=0$ in the large $\delta$ regime.
}
\end{figure}

For $0.02\lesssim\delta$ a plateau at $1/3$ appears, which widens with $\delta$.
At this plateau we observe two different magnetic orders depending of the sign of $J^{eff}_1$.
For $0.02\lesssim\delta\lesssim 0.06$, the $J^{eff}_2$ coupling  dominates and  $J^{eff}_1$ is positive;
we find a $9$-sites star shaped unit cell containing one of the patterns shown in the third line in Fig.\ \ref{fig:spin-structure-3}:
there are clear spin orientations for sites $1$ and $2$, but not for sites $3$ which are completely or partially random
(random orientation is represented by a purple square).
For $0.06\lesssim\delta$,  $J^{eff}_2$ and $J^{eff}_3$ dominate and $J^{eff}_1$ turns negative.
The transition is signaled by a clear widening of the plateau in Fig.\ \ref{fig:phasediagram_magx}.
The consequence is that the magnetic configurations no longer satisfy the Kagome ice rules. At this regime a $18$-sites magnetic unit cell is completely determined, as shown in the first line in Fig.\ \ref{fig:spin-structure-4}.

For the $\langle M_x\rangle=0$ plateau we have a similar situation depending on the value of $\delta$.
the transition around $\delta\simeq 0.06$, where $J^{eff}_1=0$, is noticed as a narrowing of the plateau.
Again, for $\delta\lesssim 0.06$ there are two possible $9$-sites star shaped magnetic unit cells,
sketched in the bottom line in Fig.\ \ref{fig:spin-structure-3}. For $0.06\lesssim \delta$  (negative $J^{eff}_1$) the magnetic unit cell consists of  $12$ sites with two possible arrangements,
as shown in Fig.\ \ref{fig:spin-structure-4}. From these cells, the magnetic tiling is obtained by translation along the $\{2\vec{a},2\vec{b}\}$.

Finally, in the saturation configuration (Fig.\  \ref{fig:spin-structure-3}),
spins at sites $1$ and $2$ are aligned with the magnetic field ({\em i.e.} $\sigma_1=+1$ and $\sigma_2=-1$).
The remaining sites $3$​​ form a triangular lattice with anisotropic couplings,
$J^{eff}_{3d}$ along horizontal bonds and $J^{eff}_3$ along the others.
The observed magnetic orderings in this sublattice, shown in Fig.\ \ref{fig:spin-structure-3} (i, ii, iii) depend on $\delta$.
For $0\leq\delta<0.005$, $J^{eff}_{3}$ is negative and dominates ($|J^{eff}_{3}|>J^{eff}_{3d}$)
giving rise to a ferromagnetic ordering; the magnetic unit cell is indicated by (i).
This phase spontaneously breaks $\mathbb{Z}_2$ symmetry through a net magnetization in the $\breve{y}$ direction.
Then, for  $0.005\leq\delta<0.045$, $J^{eff}_{3d}>0$ dominates ($|J^{eff}_3|< J^{eff}_{3d}$)
producing antiferromagnetic order along horizontal bonds; the corresponding ordering is indicated by (ii).
Finally, for $\delta\geq 0.045$, $J^{eff}_3$ is positive and dominates ($0<J^{eff}_{3d}<J^{eff}_3$)
producing antiferromagnetic order along non-horizontal bonds; the magnetic pattern is marked by (iii).

\section{Conclusions}\label{conclusions}

We have studied the effects of lattice deformations on the planar Kagome ice with nearest neighbor exchange and long range dipolar interactions. 
We have integrated out the phonon degrees of freedom and we have kept the induced effective couplings up to third neighbors. Dipolar interactions have been truncated at the same order. 
On the effective Ising model we have performed Monte Carlo simulations with an external magnetic field parallel
to one of the spins ($\vec{h}=h_y\breve{y}$) or perpendicular to it ($\vec{h}=h_x\breve{x}$). 
We have found several plateaux in the magnetization curves depending on the effects of the deformations $\delta$ (which reflects the strength of the spin-phonon coupling).

In the first case and for small spin-phonon coupling plateaux at $0$ and $1/2$ of saturation appear. As this coupling increases, the $1/2$ plateau splits into a $1/3$ and a $2/3$ plateau.
This situation persists until the deformations change the sign of the nearest neighbor effective interaction, which no longer induces frustration, thus the Kagome ice rules are not obeyed. 
In this strongly coupled regime a plethora of plateaux appear. In all these plateaux different ordered structures show up which can be easily characterized by small magnetic unit cells.

In the second case, since one every three spins is decoupled from the magnetic field, the situation is much simpler. There are again plateaux at $0$ and $1/2$ for small $\delta$, and the $1/2$ plateau turns into a $1/3$ plateau as the spin-phonon coupling is increased.

In connection to the experiments that motivated the present work \cite{GrigeraExper}, one should notice that the regime of interest is that of small Kagome plane field components, $h_x$ and/or $h_y$. We find that in this regime, both phase diagrams show the same transitions from $M=0$ to $M=1/2$ for small deformation effect $\delta$ and from $M=0$ to $M=1/3$ for larger $\delta$. The relevance of these results to the experiments remains to be analyzed, in particular by considering a more realistic model including out-of plane components of Ising spins. This will be discussed elsewhere.

\begin{acknowledgments}

The authors specially thank S.~Grigera and R.~Borzi for communicating and discussing their results. This work was partially supported by CONICET (PIP 1691) and ANPCyT (PICT 1426).

\end{acknowledgments}

\bibliographystyle{science}
\bibliography{K-I-Frefs}

\end{document}